# Internet - assisted risk assessment of infectious diseases in women sexual and reproductive health

**Andrzej Jarynowski, Damian Marchewka, Andrzej Buda**

Smoluchowski Institute of Physics, Jagiellonian University, Kraków, Poland
Interdisciplinary Research Institute, Głogów/Wrocław, Poland

**ABSTRACT**

We develop open source infection risk calculators for patients and healthcare professionals as apps for hospital acquired infections (during child-delivery) and sexually transmitted infections (like HIV). Advanced versions of e-health in non-communicable diseases do not apply to epidemiology much. There is, however, no infection risk calculator in the Polish Internet so far, despite the existence of data that may be applied to create such a tool.

The algorithms involve data from Information Systems (like HIS in hospitals) and surveys by applying mathematical modelling, Bayesian inference, logistic regressions, covariance analysis and social network analysis. Finally, user may fill or import data from Information System to obtain risk assessment and test different settings to learn overall risk.

The most promising risk calculator is developed for Healthcare-associated infections in modes for patient/hospital/sanitary inspection. The most extended version for hospital epidemiologists may include many layers of hospital interactions by agent-based modeling. Simplified version of calculator is dedicated to patients that require personalized hospitalization history of pregnancy described by questions represented by quantitative and qualitative variables. Patients receive risk assessment from interactive web application with additional description about modifiable risk factors.

We also provide solution for sexually transmitted infections like HIV. The results of calculations with meaningful description and percentage chances are presented in 'real-time' to interested users. Finally, user fills the form to obtain risk assessment for given settings.

**Introduction. Risk calculators in the concept of epidemiological intelligence**

There are many possible applications for e-health and m-health in contemporary digital age. However, a little space is devoted to IT tools that contain automatic estimation of infection risks. Moreover, digitalization of health care sector in Poland is mainly limited to accounting and financial services despite the tremendous possibilities of telemedicine. Due to the intimate nature of sexuality related issues, Internet could provide intelligent solutions.

Artificial Intelligence has already influenced medicine, e.g. in the form of regressive decision algorithms and reliability of analyzes by automating processes and detecting new dependencies in structured data sets (Jarynowski, 2009). Every Google or Facebook user may receive benefits from a personal assistant and knowledge navigator based on artificial intelligence. There are no technological obstacles to the expansion of personalized medicine. E-health uses modern information and telecommunications technologies to satisfy patient's need. Any escape from digitization is not possible. Digital revolution in medicine is happening in front of our eyes. Also Polish sociological data (Jarynowski & Serafimovic, 2014) can be processed from national polls (IPSOS, 2015). Therefore, English-speaking applications are not popular in Poland because of social differences. For this purpose, we work on an implementation consisting of:
- estimating the risk of infection on an individual basis (i.e a web application that may compute the probability of infection itself).
- e-diagnosis (symptom-based disease classification) represented by a mobile application that facilitates pre-medical identification of an infection.
- self-reporting by the National Sanitary Inspection system that may gather individual reporting of probable infections in order to bypass the clinic as an official stakeholder.

It should be noted that we consciously omit the final element (Jędrychowski, 2002) in the epidemic chain (from risk, through diagnosis to reporting and treatment) and we do not intend to support self-treatment due to the antibiotic resistance and protection of the drug itself. It is already known that a computer-aided treatment planning can be more efficient and safer than "analog" with the involvement of the infectious physician (Leibovici, et al., 2013). However, practically every drug or its replacement can be ordered externally via Internet, from outside the European Union, and the products for sexual dysfunction may be supplied according to a main offer by specialized intermediary companies.

**E-health and context of Poles**

The sexual life of Poles, despite considerable media interest, is still hardly understood. From epidemiologic point of view, we do not know much about the health of Poles (WHO has classified sexually transmitted diseases in Poland at the level of third world countries). Knowledge about (Stec, 2015) STI / STDs (sexually transmitted infections / diseases) in the Polish society is not so high, and the majority of associations may appear to be related to HIV. Other diseases seem to be an outdated issue of the past. Prophylactic programs or sex/health education are often based on non-genetic stereotypes and do not prepare for new waves of endemic pathogens in neighboring countries. An example of the propagation of non-scientific "propaganda" (Study Safe, CM-UJ Krakow 2012) is an information leaflet issued on the occasion of the spring day by the highest ranked Polish university with the longest medical tradition. In this leaflet, it appears that "condoms do not protect against HIV" (although they reduce the probability of transmission by more than hundred times), and other infections (which are thousands of times more common in Poland) are excluded. Since change we observe in sexual behavior (the age of initiation decreases and the number of partners increases (Lew-Starowicz, 2010)), the risks associated with sexually transmitted infections will not disappear, despite the increasingly effective treatment, especially the virulence of people with HIV. In Poland, there are two types of sexual education: the first one is education for cleanliness and abstinence (without information about contraception), the second one focuses on denial of all potentially discriminatory activities (Mijas, et al., 2016). Our aim is to make user become aware of risk [Fig. 1] or risk reduction methods, and choices with its consequences.

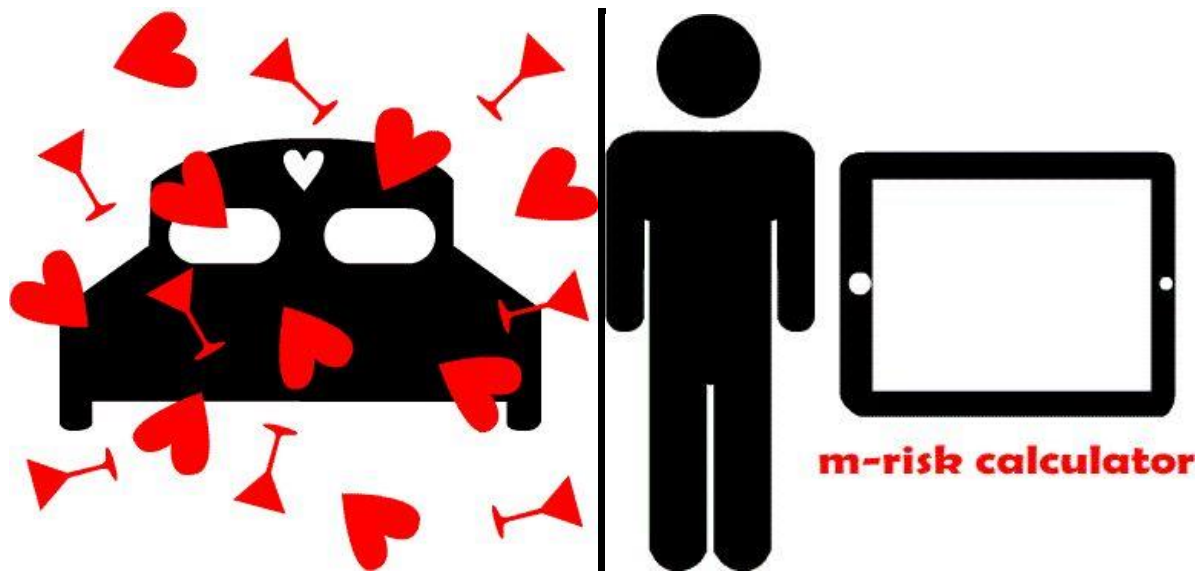

Fig.1.

Infection risk calculator of sexually transmitted infections – a possible usage schema [own pictures]

Next-generation electronic resources, including the Internet, provide the opportunity to search for information at lightning speed (Di Sia, 2015). Actually, you can find information about symptoms of common diseases and diagnose yourself. Public information ("Google Doctor" web search, medical encyclopedias, a huge number of thematic portals or social media forums) could made people become impatient and demanding. Symptoms of illness are one of the most popular topics on the Internet (what are my medicines, leaflets, information and opinions about doctors, natural medicine (Stec, 2015)). Users may apply the Internet websites for medical purposes because of the expected benefit and favorable circumstances (Szmigielska, et. al, 2012). Despite the commitment of society and the vast variety of the pro-health applications, there are no products of "epidemiological" intelligence available on the Polish Internet. In spite of everything, we can observe some trends and one may learn more and more about health by using mobile phones or computers. At least 99% of young Poles are Internet users. Poles increasingly appreciate all the benefits, for example, the possibility of obtaining advice without any real appointment. Moreover, the digital awareness of Polish women regarding the use of on-line advice is significantly higher than men that use online consulting.

Therefore, young Poles, who are no different from their Western European counterparts, may have an opportunity to become Digital Agents of Change, which disseminate digital competences in Polish society (Pokojska, 2015). Young Poles (age between 16 and 24) declare the highest frequency of computer use throughout the European Union, especially in the area of consulting/ counselling services. In the area of sexual health, women take care about their men more than men care about themselves (Izdebski, 2012). Thus, women may use our systems more likely. Moreover, healthy pregnant women must face hospital and risk of hospital infection during child-delivery. Hospitals are risky places for everybody because of various diseases and pathogens that are very likely to occur because hosts' immune system (both women and child) is weak because of delivery. Our goal is to create a mobile application [Fig. 1] that will take control on the probability of infections for different pathogens [Fig. 2]. Our program will be free to all the users, giving risk of infection at any time:

- for a single sexual act on the basis of known or implied partner attributes and sexual act characteristics;
- for a child delivery on the basis of known or implied medical circumstances.

Pregnant women and parents with babies are among the over-average people involved in the virtual world. In addition, this is a social category strongly focused on the consumption of goods, including virtual services. Hence, a number of e-health services like pregnancy monitors or diet programs can be found on the Internet already. To provide a risk calculator for future potential births, we created an online questionnaire designed for deterministic algorithms and maps of relationships between various risk factors (Jarynowski & Liljeros, 2015). We ask to provide us a fair answer up to the best knowledge. The questionnaire contains 3 groups of questions about A) women health; B) type of childbirth; C) terms of delivery D) optional - conditions of CS (Cesarean section) operation.

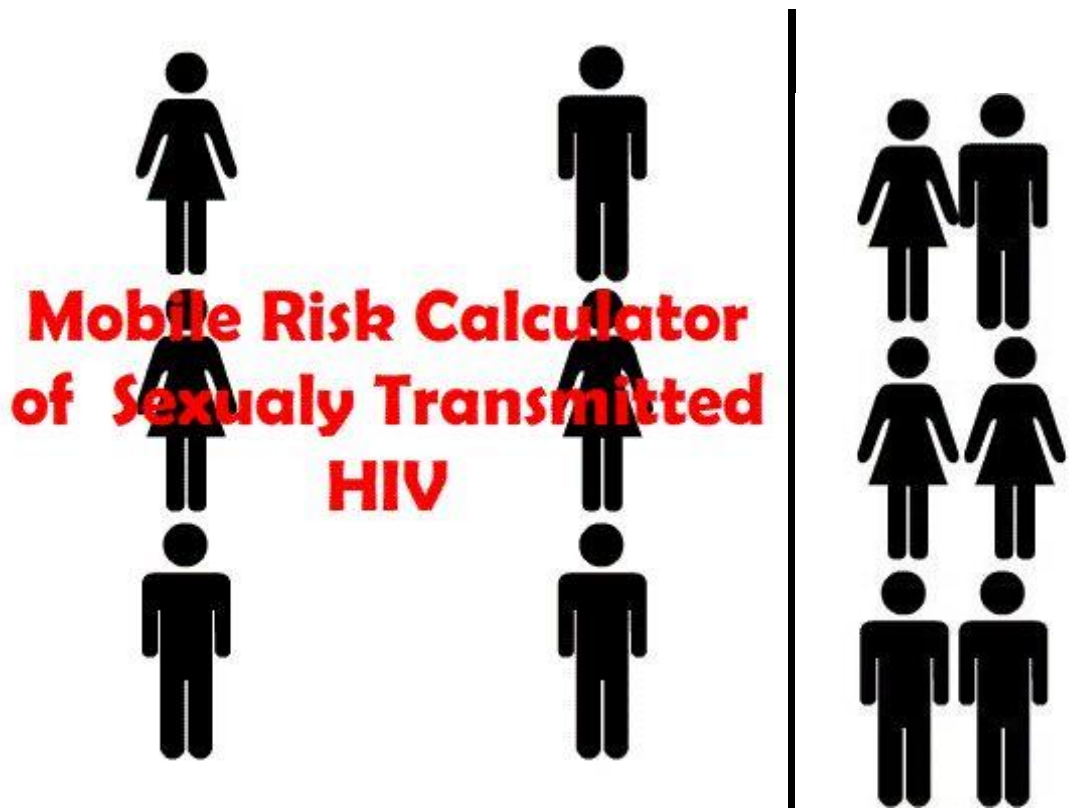

Fig.2.

Infection risk calculator of sexually transmitted infections – a possible coupling schema [own pictures]

After filling in a brief online survey, users will maintain numerical probabilities based on our algorithm. The questions are adopted from the literature on risk factors. As far as we are concerned, some effects have to be entered manually and the model produces correct results in the order of magnitude. However, we intend to confirm each factor with the quotation parameter. Vocabulary as a content of questions is also in the pilot phase and students intend to explore the perception of cognitive categories that we ask for.

**Apps and their Algorithms**

Our project is an example of the modern e-health and m-health (health in mobile phones). Epidemiological data on HIV risk have been already modelled mathematically and some services are available on the Internet. However, in literature or in epidemiological experiences we have not found any IT tool to quantify the risk of infection with an uncertain partner, but only qualitative identifications of risk groups. Perception of risk is also another phenomenon that must be processed carefully (Kahneman, 2013). Due to the very delicate nature of sexuality, we are aware of potential difficulties. Despite many controversies and serious weight of this problem, it is worth addressing this issue.

Fig.3.

Infection risk calculator of sexually transmitted infections – part of questionnaire [own pictures]

The algorithm based on the response to the sexual act (information about the type of contact- $I_f$) and the person with whom this relationship occurs (contact person information – $I_h$). These parameters determine the probability of Z-infection:

$P(Z) = P(H \setminus I_h) * P(F \setminus I_f)$.

It may be written as the probability product: chance of contact with the infected (H - event that a person is infected) and probability of infection during contact with an infected (F - event that a pathogen would be transmitted).

Fig.4.

Sexually transmitted infections calculator – English translation of part or questionnaire from Fig. 3 (for illustration purpose only, so only Polish version was implemented)[own pictures]

In child-delivering app, the most important field in the questionnaire is the question about childbirth (Caesarean section or natural) because this choice determines the risk strongly. Technically, it has been resolved as a checkbox. When we ask about perinatal circumstances we add another possible answer "do not know", because respondent may not be aware of self-medical state. Therefore, technically this section was implemented in the form of a checkbox. The last part of the questionnaire deals with operational circumstances and is performed by patients testing for cesarean section. Technically this was implemented as a Checkbox with one slider [Fig. 5].

Fig. 5.
Infection risk calculator of hospital infections during childbirth – part of questionnaire [own pictures]

The logistic regression is a mathematical model which allows us to describe the influence of many variables $X = (X_1, ..., X_i, ...)$ on the probability of infection $P(Y)$. The regression coefficients *b* and the state of the independent variables *X* by the function *Y* state give an individual the risk estimation for the patient being examined.

$$Y=\text{Rand}() + \Sigma b_i * X_i + \Sigma b_{i,j} * X_i * X_j$$

Fig. 6.

Infection risk calculator of hospital – English translation of part or questionnaire from Fig. 5 (for illustration purpose only, so only Polish version was implemented)[own pictures]

We are currently validating our 'computer assisted' risk assessment with 'human' risk assessment in a prospective study for both calculators to see how risk is understood by users [Fig. 7].

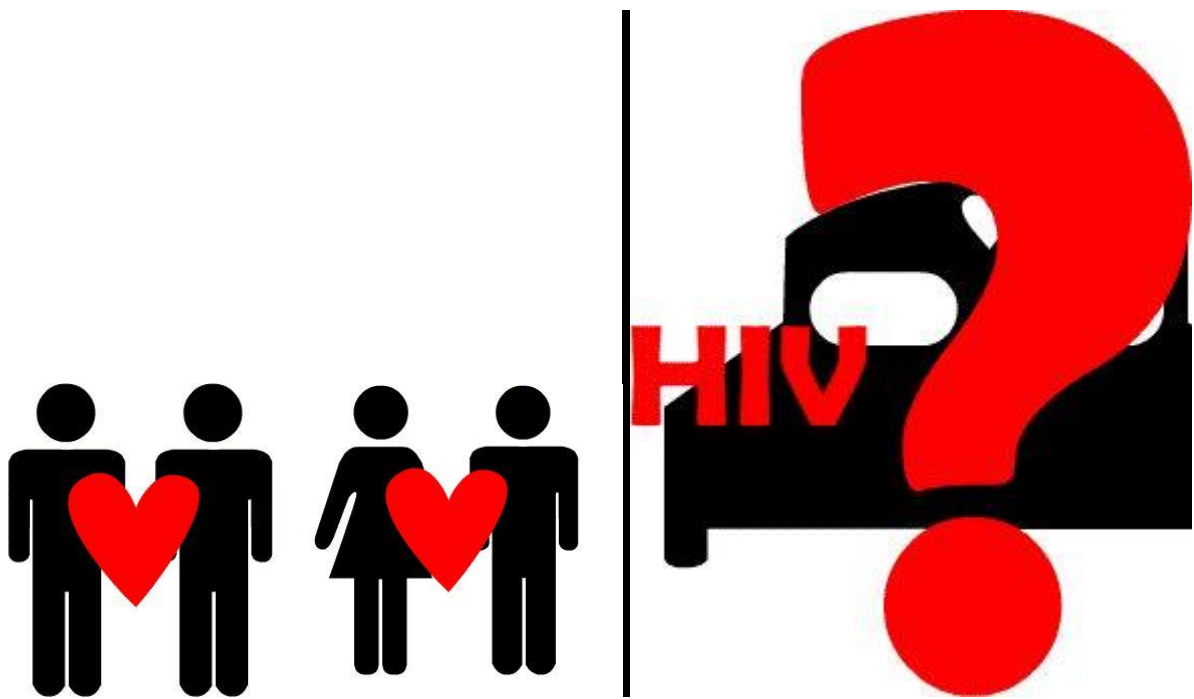

Fig. 7.

Infection risk calculator of sexually transmitted infections – How much do you risk? [own pictures]

**Conlusions**

Digital epidemiology (technique developed in this project) is also a future science, because the ability to analyze large amounts of data (Big Data) at low cost may optimize medical processes. We live in an information society where individual medical data can be used to improve epidemiological safety (Jarynowski & Grabowski, 2015). In this situation, support for sexual and reproductive health decision-making is possible already. In the age of information technology, the support of these processes by intelligent computer systems seems to be the best solution to this problem. Information technology provides tools to control infections in the individual and public health contexts (Camitz, 2010). The Internet and the associated development of IT methods may provide a number of tools for fight against infectious diseases including hospital infection and HIV. It is based on the analysis of data from widely understood dating portals (Jarynowski, 2013) and modern online sampling techniques (Lu, 2013) that boosted a huge increase in knowledge about human sexuality at the beginning of the 21st century. Internet offers a wide range of services, more than fast way Internet searches. In spite of all, the idea of developing the participation of patients in its treatment process should be carefully considered because access to existing technology is still limited.

The algorithms that we developed are a subject to criticism from the clinician community (representing medical decision makers) that do not accept the democratization of e-solutions in the patient-doctor relationship. On the other hand, the market (represented mainly by pharmaceutical companies) that creates various types of calculators itself, is not interested in commercializing of such solutions. In conclusion, we think our HIV calculator could complement other "party" or "alcoholic" apps [Fig. 7]. The Childbirth calculator may be a part of "pregnancy" apps. Proposed "gamification" that makes user guess the risk does not provide more knowledge than giving a system in which everyone can acquire, with a large degree of autonomy, competence for initiating processes of exploitation of own creative personality (Di Sia, 2016).

Acknowledgment

We want to thank e-methodology society for inspiration. AJ and DM were partly supported by NCBiR founds for Social Innovations. Links to apps:

- Child delivery http://platforma.sirsz.pl/ankieta/zak/index.php

- HIV http://interdisciplinaryresearch.eu/index.php/ankieta/

**REFERENCES**

Camitz, M. (2010). Computer Aided Infectious Disease Epidemiology - Bridging to Public Health, PhD thesis, Karolinska Institutet, Stockholm.

Di Sia, P. (2015). About Internet and the Diffusion of Science, *E-methodology*, *2*, 18-26

Di Sia, P. (2016). The Internet and new educational perspectives, *E-methodology*, *3*, 18-24

IPSOS (2015). Diagnoza stanu wiedzy w zakresie HIV/AIDS oraz zakażeń przenoszonych drogą płciową (ZPDP) [Diagnosis of the State of Knowledge on HIV/AIDS and Sexually Transmitted Infections].

Izdebski, Z. (2012). *Seksualność Polaków*, Warszawa: MUZA SA [Sexuality of Poles].

Jarynowski, A. (2009). Wirtualne aspekty nauki i techniki. Retrieved from http://www.racjonalista.pl/kk.php/s,6648. [Virtual Aspects of Science and Technology]

Jarynowski, A. (2013). Modelowanie epidemiologiczne na sieciach społecznych na przykładzie zakażeń szpitalnych (HAI) i chorób przenoszonych drogą płciową (STI). Studia i Materiały Informatyki Stosowanej10 (13) [Epidemological Modelling on Social Networks on the Example of Hospital Acquired Infection (HAI) and Sexually Transmited Infections (STI)]

Jarynowski, A., & Serafimovic, A. (2014). Studying possible outcomes in a model of sexually transmitted virus (HPV) causing cervical cancer for Poland. Advances in Intelligent System and Computing, 229 (2)

Jarynowski, A., & Grabowski, A. (2015). Modelowanie epidemiologiczne dedykowane Polsce. Portal CZM 9(6). [Epidemiological Modelling Dedicated to Poland]

Jarynowski, A., & Liljeros, F. (2015). Contact networks and the spread of MRSA in Stockholm hospitals. In Network Intelligence Conference (ENIC), 2015 Second European (pp. 150-154). IEEE

Jędrychowski, W. (2002) *Podstawy Epidemiologii. Metody badań oraz materiały ćwiczeniowe*. Kraków: Wydawnictwo Uniwersytetu Jagiellońskiego. [Foundations of

Epiedmiology. Methods of Research and Exercise Materials]

Kahneman, D. (2003). A perspective on judgment and choice: mapping bounded rationality. *American psychologist*, *58(9)*, 697.

Leibovici, L., Kariv, G., & Paul, M. (2013). Long-term survival in patients included in a randomized controlled trial of TREAT, a decision support system for antibiotic treatment. *Journal of Antimicrobial Chemotherapy*, *68(11)*, 2664-2666

Lew-Starowicz, Z (2010). *Podstawy Seksuologii*. Warszawa: PZWL [Fundations of Sexuology]

Lu, X. (2013). Respondent-driven sampling: theory, limitations & improvements. PhD thesis, Karolinska Institutet, Stockholm.

Mijas, M., Dora, M., Brodzikowska, M., & Żołądek, S. (2016). *Język, media, HIV: obraz zakażenia i osób seropozytywnych w artykułach prasowych*. Kraków : Stowarzyszenie Profilaktyki i Wsparcia w zakresie HIV/AIDS "Jeden Świat". [Language, Media, HIV: Image of Infection and Seropositive Persons in Press Articles]

Pokojska, J. (2015). Agentki cyfrowej zmiany – kompetencje cyfrowe kobiet w Polsce. Report DeLab UW. [Agents of Digital Change – Digital Competences of Women in Poland].

Study Safe – StudjUJ bezpiecznie (2012). Kraków: Wydawnictwo Uniwersytetu Jagiellońskiego

Stec, M. (2015). Postawy i potrzeby informacyjno-edukacyjne w kontekście zakażeń przenoszonych drogą płciową, w tym HIV/AIDS. MILLWARD BROWN Report. [Attitudes and Information-Educational Needs in the Context of Sexually Transmitted Infections, Including HIV/AIDS]

Szmigielska, B., Wolski, K., & Jaszczak, A. (2012). Modele wyjaśniające zachowania użytkowników. E-mentor, 3 (45). internetu [Models Explaining Behavior of Internet Users]